\documentstyle[preprint,aps,tighten]{revtex}

\begin{document}
\title{Diquark-diquark correlations in the $^1S_0$ $\Lambda \Lambda$
potential}
\author{T. Fern\'andez-Caram\'es$^{(1)}$, A. Valcarce$^{(2)}$, and P. Gonz\'a%
lez$^{(1)}$}
\address{$(1)$ Dpto. de F\' \i sica Te\'orica and IFIC\\
Universidad de Valencia - CSIC, E-46100 Burjassot, Valencia, Spain}
\address{$(2)$ Grupo de F\' \i sica Nuclear and IUFFyM\\
Universidad de Salamanca, E-37008 Salamanca, Spain}
\maketitle

\begin{abstract}
We derive a $\Lambda \Lambda$ potential from a chiral
constituent quark model that has been successful in describing one, two and
three nonstrange baryon systems. The resulting interaction at low energy is
attractive at all distances due to the $\sigma $ exchange term. The
attraction allows for a slightly bound state just below the $\Lambda \Lambda 
$ threshold. No short-range repulsive core is found. We extract the
diquark-diquark contribution that turns out to be the most attractive and
probable at small distances. At large distances the asymptotic behavior of
the $\Lambda \Lambda $ interaction provides a prediction
for the $\sigma \Lambda \Lambda $ coupling constant.
\end{abstract}

\vspace*{2cm}

\noindent Keywords: diquarks, chiral constituent quark models, $\Lambda
\Lambda$ interaction \newline
\noindent Pacs: 12.39.Jh, 12.39.Pn, 14.20.Pt

\newpage

The $\Lambda \Lambda $ interaction has been the object of extensive
consideration. On the one hand the knowledge of the $%
\Lambda \Lambda $ altogether with the $N\Lambda $ interaction is a necessary
ingredient to obtain a physical description of hypernuclei for which binding
energies have been measured \cite{Exp00}. On the other hand the $\Lambda
\Lambda $ $^{1}S_{0}$ channel has the quantum numbers of the $H-$dibaryon,
theoretically predicted almost thirty years ago with a mass below the $%
\Lambda \Lambda $ threshold \cite{Jaf77} and not experimentally identified
until now \cite{Exp01}.

Since $\Lambda $ targets are not available there is not direct measurement
of the $\Lambda \Lambda $ scattering process. Indirect information can be
extracted from the scarce old $\Lambda p$ scattering data \cite{AKS} and
from measured binding energies of $\Lambda \Lambda $ hypernuclei \cite{Exp00}%
. Approaches to the $\Lambda \Lambda $ interaction potential have been
pursued at the baryon and quark levels. At the baryon level phenomenological
(in the sense of reproducing some indirect experimental information)
meson-exchange ($\sigma $ and $\omega )$ potentials have been constructed 
\cite{Rij99,Reu94,Car99}. At the quark level, gluon dominant (only one-gluon
exchange between quarks), pure chiral (only meson exchanges between quarks)
and chiral constituent quark (gluon plus meson exchanges) models have been used \cite%
{SSY}. 
Finally, hybrid models where some pieces of the interaction (usually the
gluon and pion exchanges) are considered at the quark level while others
(usually the one-sigma exchange) are parametrized at the baryonic level\cite%
{Str88}, with several free parameters fitted differently for
different observables (deuteron binding energy, $NN$ singlet $S-$wave phase
shifts, $NN$ $P-$wave phase shifts and hyperon-nucleon interaction), have been
also efficient to describe data.

A quite general conclusion arising from the quark approaches until now is that the
interaction between two $\Lambda ^{\prime }s$ presents a repulsive core \cite{SSY}.
Though this conclusion is easily understood in a gluon dominant model by
comparison to the $NN$ case (the interaction generated at short distances by
the one-gluon exchange is repulsive) it seems more controversial in
chiral constituent quark models incorporating the $\sigma -$meson exchange,
responsible for the correct medium range behavior in the $NN$ system and
attractive at all distances, where it may be dominant.

Comparatively phenomenological baryonic interactions can be very
precise when a sufficient amount of data are at disposal (the interaction
parameters depending on the specific case under consideration), however quark
treatments can be fully predictive since the same potential parameters
(fixed once for all, for instance, in the well-known nonstrange sector)
could be applied to any other baryon-baryon system.
To this respect we should
first realize that at the quark level only a chiral constituent quark model
can provide a realistic description of nonstrange two-baryon systems.
The gluon dominant model is not able to reproduce the long-distance behavior
of the $NN$ interaction. The pure chiral models lack, when applied to the $%
NN $ interaction, of the necessary medium range attraction \cite{Nak98}.
Then we are left with chiral constituent quark models where the appearance
or not of a repulsive core in the $\Lambda \Lambda $ interaction will depend
on the interplay between gluon and meson exchanges, i.e., on the relative
intensity of the one gluon and the meson exchange terms in the potential.
This relative intensity can be fixed, once a chiral realization of QCD at
low energy is chosen, by the requirement of having the most accurate
description of one and two-body systems.

In this letter we revise the conclusions obtained about the $\Lambda \Lambda$
system by means of a chiral constituent quark model tightly constrained by the
description of the nonstrange sector and the strange meson and baryon spectra.

In the last years a high degree of precision has been reached in the
description of the nonstrange one, two and three baryon systems by means of
a $SU(2)\times SU(2)$ chiral constituent quark model (as originally
named in reference \cite{Fer93}) containing $\pi $ and $\sigma $ exchanges plus a
residual one-gluon exchange interaction \cite{Val05}. We shall call this
model hereforth CCM \cite{Fer93}. This strong interaction model corresponds 
to a nonrelativistic approach to an effective theory incorporating quarks, 
gluons and Goldstone bosons \cite{Man84}, the effectiveness of the CCM 
parameters hopefully giving account, in an effective manner, of effects 
out of the scope of our approach.
A main difference with other models with a {\it similar}
content \cite{Shi99} is the consistent treatment, at the level of the wave
function, of the baryon-baryon interaction and the baryon spectra. 
Another basic difference comes from
the requirement to describe a wide number of observables, restricting very much
the possible values of the parameters. 
The predictive power of the CCM has been tested in its successful application to
the $NN,$ $N\Delta $, tritium and nonstrange baryon spectra. The model
has a reduced
relative intensity of the gluon against meson exchanges, as compared
to the other models. Gluon and pion contribute both about half of
the $\Delta -N$ mass difference and the short-range $NN$ repulsion in the $%
S- $wave comes mainly from the pion term at difference with other models
where the gluon, through configuration mixing, provides the major
contribution. This shows that {\it similar}
models, in the sense of using a similar form of
the interaction, may give rise to very
different physical descriptions.

In order to apply the CCM to the $\Lambda \Lambda $ case an extension to
deal with strange baryons is required. Besides confinement and gluon terms
the minimal chiral extension used in the literature comprises
the exchange of a singlet scalar meson, the $\sigma $ (mediating
light-strange and strange-strange as well as light-light quark interactions)
and a pseudoscalar octet involving pions, kaons and eta. The motivation for
this systematic is the following. From existing baryonic analysis the $%
\sigma $ and $\omega $ exchanges seem to be the most relevant ones. At the
quark level one expects that quark antisymmetrization effects play a similar
role to the baryonic $\omega $ exchange. Indeed at the baryon level the $%
\omega $ provides the short-range repulsion in the $NN$ interaction whilst
at the quark level antisymmetrization effects on the gluon and $\pi $ terms
provide to a great extent the same effect. The consideration of an $%
SU(3)\times SU(3)$ chiral model would incorporate an octet of scalar mesons
as well. However the scalar masses, with the exception of the singlet $%
\sigma $ $(f_{0}(600))$, are significantly higher than the pseudoscalar
octet ones what makes plausible their contributions to be less important. 
This way of proceeding results in an extended model that fits well
the strange hadron spectra \cite{Vij05} while keeping the successful
description for the nonstrange systems.

Explicitly the quark-quark interaction reads: 
\begin{equation}
V_{qq}(\vec{r}_{ij})=V_{CON}(\vec{r}_{ij})+V_{OGE}(\vec{r}_{ij}) +V_{\pi}(%
\vec{r}_{ij})+V_{\sigma}(\vec{r}_{ij}) +V_{K}(\vec{r}_{ij})+V_{\eta}(\vec{r}%
_{ij}) \,\, ,  \label{eq1}
\end{equation}
where the $i$ and $j$ indices are associated with $i$ and $j$ quarks,
respectively, and ${\vec{r}}_{ij}$ stands for the interquark distance.

The confinement potential is chosen linear, 
\begin{equation}
V_{CON}({\vec{r}}_{ij})=-a_{c}\,{\vec{\lambda}}_{i}\cdot {\vec{\lambda}}%
_{j}\,r_{ij} \,\, ,
\end{equation}%
where $a_{c}$ is the confinement strength, the ${\vec{\lambda}}$'s are the $%
SU(3)$ color matrices, and the color structure prevents from having
confining interaction between color singlets.

From the nonrelativistic reduction of the one-gluon-exchange diagram in QCD
for point-like quarks one gets, 
\begin{equation}
V_{OGE}({\vec{r}}_{ij})={\frac{1}{4}}\,\alpha _{s}\,{\vec{\lambda}}_{i}\cdot 
{\vec{\lambda}}_{j}\Biggl \lbrace{\frac{1}{r_{ij}}}- {\frac{1 }{4}} \left( {%
\frac{1}{{2\,m_{i}^{2}}}}\, + {\frac{1}{{2\,m_{j}^{2}}}}\, + {\frac{2}{3 m_i
m_j}}{\vec{\sigma}}_{i}\cdot {\vec{\sigma}}_{j} \right) \,\,{\frac{{%
e^{-r_{ij}/r_{0}}}}{{r_{0}^{2}\,\,r_{ij}}}} \Biggr \rbrace\,\, ,  \label{reg}
\end{equation}%
where $\alpha _{s}$ is an effective strong coupling constant and $m_{i}$ is
the mass of the quark $i$. ${\vec{\sigma}}_{i}$ stands for the Pauli spin
operator. Let us realize that the contact term involving a Dirac $\delta (%
\vec{r})$ that comes out in the deduction of the potential has been
regularized in the form 
\begin{equation}
\delta (\vec{r}\,)\,\rightarrow \,{\frac{1}{{4\pi r_{0}^{2}}}}\,\,{\frac{{%
e^{-r/r_{0}}}}{r}}\,\, ,
\end{equation}%
giving rise to the second term of Eq. (\ref{reg}) ($r_{0}$ is a
regularization parameter). This avoids to get an unbound baryon spectrum
from below when solving the Schr\"{o}dinger equation \cite{Bha80}.

The static pion and sigma exchange potentials are given by: 
\begin{equation}
V_{\pi} ({\vec r}_{ij}) = {\frac{1 }{3}} {\frac{ g^2_{ch} }{{4 \pi}}} \, {%
\frac{m^2_{\pi} }{{4 m_i m_j}}} \, {\frac{\Lambda_{\chi}^2 }{%
\Lambda_{\chi}^2 - m_{\pi}^2}} \, m_{\pi} \, \left[ \, Y (m_{\pi} \, r_{ij})
- {\frac{ \Lambda_{\chi}^3 }{m_{\pi}^3}} \, Y (\Lambda_{\chi} \, r_{ij}) %
\right] {\vec \sigma}_i \cdot {\vec \sigma}_j {\vec \tau}_i \cdot {\vec \tau}%
_j \, ,  \label{OPE}
\end{equation}
\begin{equation}
V_{\sigma} ({\vec r}_{ij}) = - {\frac{ g^2_{ch} }{{4 \pi}}} \, {\frac{%
\Lambda_{\chi}^2 }{\Lambda_{\chi}^2 - m_{\sigma}^2}} \, m_{\sigma} \, \left[
Y (m_{\sigma} \, r_{ij})- {\frac{\Lambda_{\chi} }{{m_{\sigma}}}} \, Y
(\Lambda_{\chi} \, r_{ij}) \right] \, ,  \label{OSE}
\end{equation}
where $g_{ch}$ is the chiral coupling constant, $\Lambda_{\chi}$ is a
cut-off parameter and the ${\vec \tau}$'s are the isospin quark Pauli
matrices. $m_{\pi}$ and $m_\sigma$ are the masses of the pseudoscalar and
scalar Goldstone bosons, respectively. $Y(x)$ is the standard Yukawa
function defined by $Y(x)=e^{-x}/x$. The {\it new} kaon and eta potentials
read

\begin{eqnarray}
V_{K}(\vec{r}_{ij}) &=&{\frac{1 }{3}} {\frac{g_{ch}^{2}}{{4\pi }}}{\frac{%
m_{K}^{2}}{{\ 4m_{i}m_{j}}}}{\frac{\Lambda _{K}^{2}}{{\Lambda
_{K}^{2}-m_{K}^{2}}}}m_{K}\left[ Y(m_{K}\,r_{ij})-{\frac{\Lambda _{K}^{3}}{%
m_{K}^{3}}}Y(\Lambda _{K}\,r_{ij})\right] (\vec{\sigma}_{i}\cdot \vec{\sigma}%
_{j})\sum_{a=4}^{7}{(\lambda _{i}^{a}\cdot \lambda _{j}^{a})}\,,  \nonumber
\\
V_{\eta }(\vec{r}_{ij}) &=& {\frac{1 }{3}} {\frac{g_{ch}^{2}}{{4\pi }}}{%
\frac{m_{\eta }^{2}}{{\ 4m_{i}m_{j}}}}{\frac{\Lambda _{\eta }^{2}}{{\Lambda
_{\eta }^{2}-m_{\eta }^{2}}}}m_{\eta }\left[ Y(m_{\eta }\,r_{ij})-{\frac{%
\Lambda _{\eta }^{3}}{m_{\eta }^{3}}}Y(\Lambda _{\eta }\,r_{ij})\right] (%
\vec{\sigma}_{i}\cdot \vec{\sigma}_{j})\left[ cos\theta _{p}(\lambda
_{i}^{8}\cdot \lambda _{j}^{8})-sin\theta _{p}\right] \,,  \nonumber
\end{eqnarray}
where the angle $\theta _{p}$ appears as a consequence of considering the
physical $\eta $ instead the octet one.

Let us keep in mind that the values of the parameters for
confinement, gluon, pion and sigma exchanges are fixed from the nonstrange
sector \cite{Val05}. For kaon and eta exchanges the parameters are fixed
from a fit to the strange meson spectra \cite{Vij05}. 
In Table I we compile the values used. Let us note that our cutoff parameter
for the kaon and the eta is different than for sigma and pion. This comes
from a fine tune to the strange meson spectra but it will not be
quantitatively relevant for the $\Lambda \Lambda $ interaction.

The general nonrelativistic framework to obtain the $\Lambda \Lambda $
potential from the quark-quark interaction is the Resonating Group Method
(RGM). A much\ more simplified treatment, the adiabatic or Born-Oppenheimer
(BO) approximation, can be suitable under the assumption that the quarks
move inside the baryonic clusters much faster than the clusters relative to
each other. Since applications of RGM and BO to the nonstrange systems do
not show any relevant difference \cite{Gar99} we shall adhere hereforth to a BO
approach. The resulting potential can be obtained as:

\begin{equation}
V_{B_{1}B_{2}(L\,S\,T)\rightarrow B_{3}B_{4}(L^{\prime }\,S^{\prime
}\,T)}(R)={\cal V}_{L\,S\,T}^{L^{\prime }\,S^{\prime }\,T}(R)\,-\,{\cal V}%
_{L\,S\,T}^{L^{\prime }\,S^{\prime }\,T}(\infty )\,,  \label{PotBO}
\end{equation}%
where 
\begin{equation}
{\cal V}_{L\,S\,T}^{L^{\prime }\,S^{\prime }\,T}(R)\,=\,\frac{{\left\langle
\Psi _{B_{1}B_{2}}^{L^{\prime }\,S^{\prime }\,T}(\vec{R})\mid
\sum_{i<j=1}^{6}V_{qq}({\vec{r}_{ij}})\mid \Psi _{B_{3}B_{4}}^{L\,S\,T}(\vec{%
R})\right\rangle }}{\sqrt{\left\langle \Psi _{B_{1}B_{2}}^{L^{\prime
}\,S^{\prime }\,T}(\vec{R})\mid \Psi _{B_{1}B_{2}}^{L^{\prime }\,S^{\prime
}\,T}(\vec{R})\right\rangle }\sqrt{\left\langle \Psi _{B_{3}B_{4}}^{L\,S\,T}(%
\vec{R})\mid \Psi _{B_{3}B_{4}}^{L\,S\,T}(\vec{R})\right\rangle }}\,,
\label{BODEF}
\end{equation}
with $\Psi _{B_{i}B_{j}}^{L\,S\,T}(\vec{R})$ given in general ($B_{i}$ and $%
B_{j}$ can be nonidentical baryons) by 
\begin{eqnarray}
\Psi _{B_{1}B_{2}}^{LST}({\vec{R}}) &=&{\frac{{\cal A}}{\sqrt{1+\delta
_{B_{1}B_{2}}}}}\sqrt{\frac{1}{2}}\Biggr\{\left[ \Phi _{B_{1}}\left( 123;{-{%
\frac{{\vec{R}}}{2}}}\right) \Phi _{B_{2}}\left( 456;{\frac{{\vec{R}}}{2}}%
\right) \right] _{LST}  \nonumber \\
&+&(-1)^{f}\,\left[ \Phi _{B_{2}}\left( 123;{-{\frac{{\vec{R}}}{2}}}\right)
\Phi _{B_{1}}\left( 456;{\frac{{\vec{R}}}{2}}\right) \right] _{LST}\Biggr \}%
\,,  \label{Gor}
\end{eqnarray}%
where $f$ (even or odd) determines the baryon-baryon symmetry and $S$, $T$
and $L$ correspond to the total spin, isospin and orbital angular momentum
of the two-baryon system. ${\cal A}$ is the six-quark antisymmetrizer
written as \cite{Hol84} 
\begin{equation}
{\cal A} = \left ( 1-\sum_{i=1}^2\sum_{j=4}^5 P_{ij} - P_{36} \right ) (1-%
{\cal P}) \, ,
\end{equation}
where we have identified quarks 3 and 6 as the strange ones in each
baryon. $P_{ij}$ is the operator that exchanges particles $i$ and $j$ and $%
{\cal P}$ exchanges the two clusters. $P_{ij}$ can be explicitly written as
the product of permutation operators in color ($C$), spin-isospin ($ST$) and
orbital ($O$) spaces, 
\begin{equation}
P_{ij} = P^C_{ij} \, P^{ST}_{ij} \, P^O_{ij} \, .
\end{equation}
The subtraction of ${\cal V}_{L\,S\,T}^{L^{\prime }\,S^{\prime }\,T}(\infty)$
assures that no internal cluster energies enter in the baryon-baryon
interacting potential.

The total wave function of a single baryon cannot be in the $\Lambda\Lambda$
case, at difference with the $NN$ one, separated in orbital, spin and color
parts. Whereas one can isolate the color part one has to maintain the
orbital and flavor structure altogether due to the flavor (mass) dependence
of the orbital part. Hence the $\Lambda$ wave function reads 
\begin{equation}
\Phi _{\Lambda }(\vec{r_{1}},\vec{r_{2}},\vec{r_{3}};\vec{R}/2)=\frac{1}{%
\sqrt{2}}\left( \phi ^{MS}\chi ^{MS}+\phi ^{MA}\chi ^{MA}\right) \otimes \xi
\, ,  \label{wfu}
\end{equation}
$\phi$, $\chi$, and $\xi$ standing for the orbital-flavor, spin and color
parts, respectively. The orbital wave function for a quark $i$ in a cluster
at position $\vec{R}/2 $ is chosen as 
\begin{equation}
\varphi (\vec{r_{i}},\vec{R}/2)=\left( \frac{1}{\pi b_{i}^{2}}\right)
^{3/4}exp\left( -\frac{\left( \vec{r_{i}}-\vec{R}/2\right) ^{2}}{2b_{i}^{2}}%
\right) \, .
\end{equation}
The dependence on the quark mass enters in the parameter $b_{i}$. We shall
assume $1/(m_{s}b_{s}^{2})=1/(m_{u,d}b_{u,d}^{2})$ in order to have the same
kinetic energy for all quarks in the baryon.

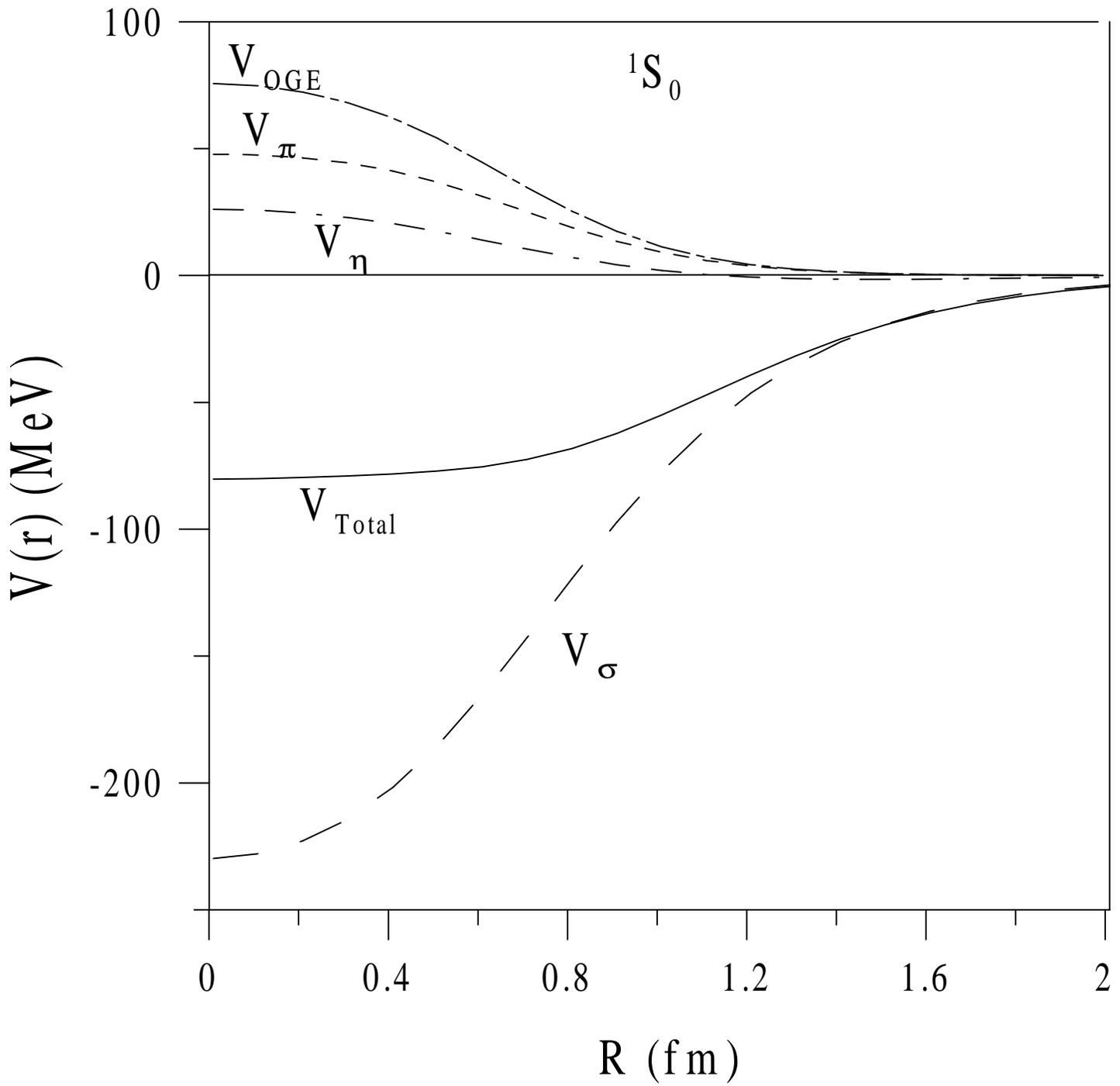
\begin{figure}[tbp]
\caption{$^1S_0$ $\Lambda \Lambda$ potential. The contribution of the
different terms in Eq. (\protect\ref{eq1}) has been depicted.}
\label{fig1}
\end{figure}

The result for the $^{1}S_{0}$ potential is drawn in Fig. \ref{fig1} where
the contributions from $\sigma$, $\pi$, $\eta$ and gluon exchanges are also
depicted. Note that the kaon contribution vanishes due to isospin
conservation in the possible quark-kaon-quark vertices. As can be seen the
dominant contribution is attractive and comes from the $\sigma$. Actually it
corresponds to a pure $\sigma$ baryonic potential since no quark-exchanges
(related to the $P_{kl}$ terms in the antisymmetrizer) contribute either to
the numerator or denominator in (\ref{BODEF}). Let us realize that except
for the small differences coming from the exchange contribution in the $NN$
case and the strange quark mass in the $\Lambda\Lambda$ one, this
interaction should be, as in fact it is, quantitatively very similar in the $%
NN$ and $\Lambda\Lambda$ cases.

The gluon contribution comes only from quark exchanges, since being the gluon
a color octet cannot directly connect two baryons (color singlet states).
Furthermore the only surviving gluonic contribution comes from the
chromomagnetic part being repulsive. The strength of this repulsion is
smaller (about one third for $R=0.01$ fm) than the corresponding one in $NN$%
. This is due, on the one hand, to the lesser number of diagrams
contributing (no exchanges involving light-strange quarks are possible) and,
on the other hand, to the strange quark mass factors in the OGE potential.
Despite this we can immediately understand why the $\Lambda \Lambda $
interaction has a repulsive core in other chiral constituent quark models
where the gluon term is dominant. As a matter of fact if we multiply our
gluon coupling constant by a factor 2 and rearrange correspondingly the chiral
coupling constant, a repulsive core would be obtained (at the prize of
destroying the good description of many other observables).

Regarding the $\pi $ term it is repulsive as in $NN$. However as for $%
\Lambda \Lambda $ it only contributes, due to its isovector character,
through quark exchanges, the repulsion is much weaker (about one fifth for $%
R=0.01$ fm). Indeed, in the $\Lambda\Lambda$ case the pion contribution
turns out to be less important than the gluon one and much less important in
absolute value than the $\sigma$ one. Hence the attraction
provided by the $\sigma -$exchange surpasses the gluon+pion repulsion. This
net attraction cannot be compensated by the small repulsive contribution
from the $\eta -$exchange. To this respect it is also interesting to look at
the results for $N\Lambda \rightarrow N\Lambda$ (see below).

Since the values of the parameters $m_{s}$ and $\Lambda _{\eta ,K}$ are fitted
only from spectroscopy it is convenient to test the sensitivity of our
results to changes in them. As mentioned before the effect of taking $%
\Lambda _{\eta ,K}$ different from $\Lambda _{\chi }$ is small, about 5\% in
the values of the potential at most. Regarding $m_{s}$ a variation of $150$
MeV around the chosen value gives rise to a modification of
the potential of a 15\% at most. This gives us confidence in the results we
obtain whose qualitative character is essentially determined by the well
fitted $SU(2)\times SU(2)$ parameters.

Once we have determined the $\Lambda \Lambda $ interacting potential we
proceed to the study of bound states. For this purpose the method of the Fredholm
determinant is particularly simple and trustable: starting from the
Lippmann-Schwinger equation for the $\Lambda \Lambda $ system, 
\begin{equation}
\int d^{3}k^{\prime \prime }\left[ \delta (\vec{k^{\prime }}-\vec{k^{\prime
\prime }})-\frac{\left\langle \vec{k^{\prime }}\left\vert V\right\vert \vec{%
k^{\prime \prime }}\right\rangle }{E-E_{k^{\prime \prime }}}\right]
\left\langle \vec{k^{\prime \prime }}\left\vert T(E)\right\vert \vec{k}%
\right\rangle =\left\langle \vec{k^{\prime }}\left\vert V\right\vert \vec{k}%
\right\rangle \, ,
\end{equation}
where $\mid \vec{k}>$ stands for a momentum state of $\Lambda \Lambda$, and
substituting the integral by a $N-$point quadrature we can formally write: 
\begin{equation}
\left[ T(E)\right] =\frac{\left[ V\right] }{\left[ 1-VG_{0}(E)\right] } \, ,
\end{equation}
where $G_{0}(E)=1/(E-\vec{p}^{\, 2}/2\mu )$ ($\mu $ is the reduced mass of
the system) is the nonrelativistic propagator. Then a bound state,
corresponding to a pole of the $T$-matrix in the real axis, leads to 
\begin{equation}
\det \left[ 1-VG_{0}(E)\right] =0 \, ,
\end{equation}%
whose solution determines the binding energy \cite{Val95}.

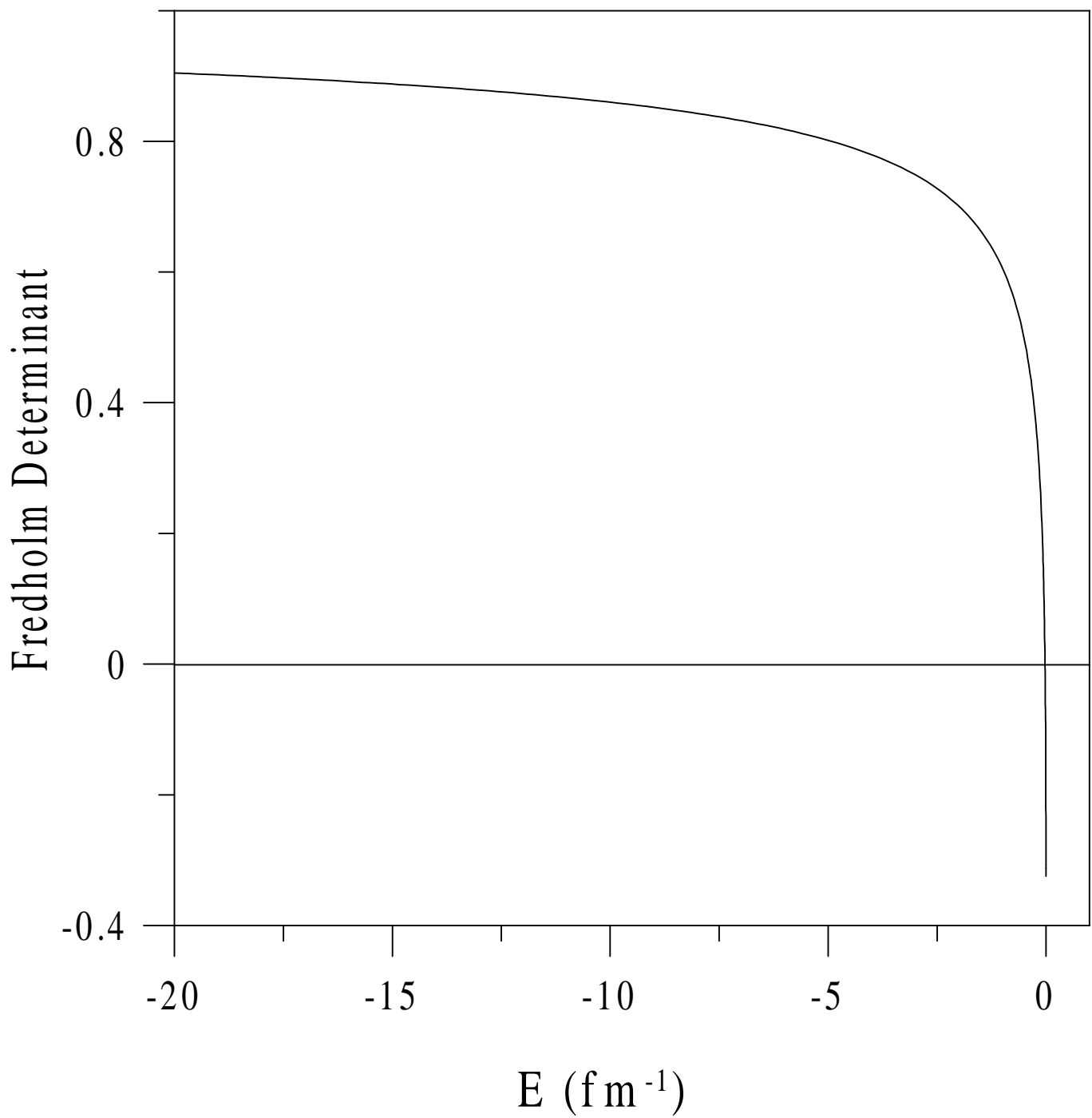
\begin{figure}[tbp]
\caption{Fredholm determinat of the $\Lambda\Lambda$ system with $J=0$ as a
function of the nonrelativistic energy E.}
\label{fig2}
\end{figure}

In Fig. \ref{fig2} the value of the Fredholm determinant as a function of
energy is depicted. As we can see there appears only one possible bound
state very close to the $\Lambda \Lambda $ threshold (with a binding energy
of 0.022 MeV). This is in agreement with upper bounds extracted from $\Lambda
\Lambda $ hypernuclei \cite{Exp00} but in contradiction with previous quark
model results obtaining much more binding what was associated to the
presence of a tightly bound $H-$dibaryon \cite{SSY}.
Although to analyze  the $H-$dibaryon a coupled channel
calculation should be performed, when this is done, the probabilities
of the  $\Lambda\Lambda$, $N\Xi$ and $\Sigma\Sigma$ components
do not differ much from the initial flavor
$SU(3)$ Clebsch-Gordan coefficients \cite{SSY,Str88}. This comes from the fact
that the coupling between the $\Lambda\Lambda$, $N\Xi$ and
$\Sigma \Sigma$ is very  small,  all the  direct  diagrams  do  not
contribute  (except for the pion between $\Lambda\Lambda$  and
$\Sigma\Sigma$,  but giving a small contribution  due  to  the
different symmetry of the spin wave function), the coupling
mainly generated by the exchange diagrams that are very
short-ranged as to give an important contribution to the binding.
Therefore the binding is mainly driven by the sigma-meson
exchange in each two-body channel separately.
Therefore we might infer, from our one-channel
calculation, that a tightly bound $H-$dibaryon as obtained with
{\it similar} quark models may not be justified. 
According to our result the experimental absence of such state could
be related to the well known difficulty 
to disentangle in partial wave analysis  the
existence of a resonance close to a threshold  (this  is
nowadays very much discussed regarding the nature  of  some
mesons  as could be, for example, the $X(3872)$, that  it  is
precisely in the $D^0D^{*^0}$  threshold).  

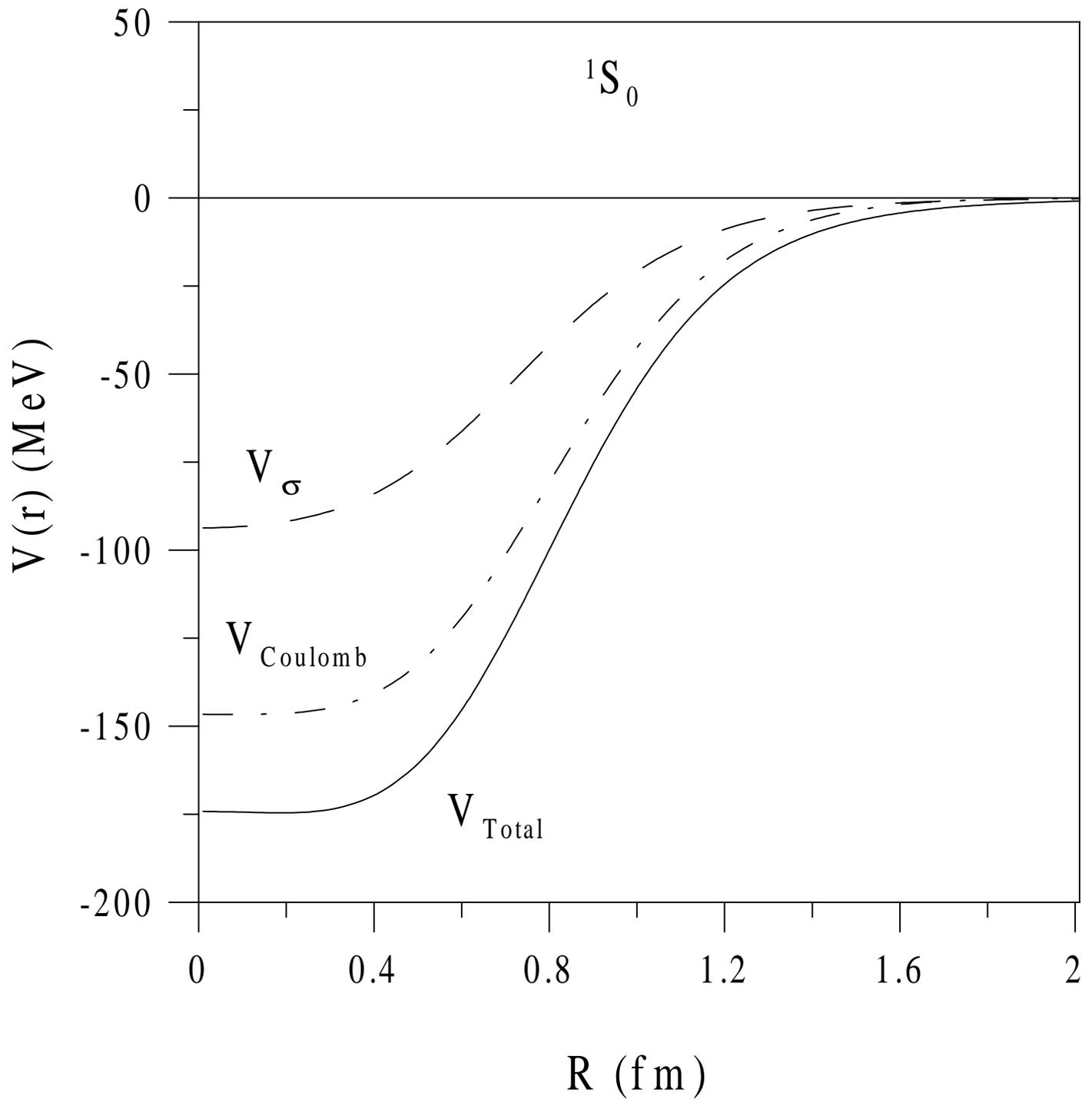
\begin{figure}[tbp]
\caption{Diquark-diquark contributions to the $^1S_0$ $\Lambda \Lambda$
potential. We have shown separately the most important terms in Eq. (\protect
\ref{eq1}).}
\label{fig3}
\end{figure}

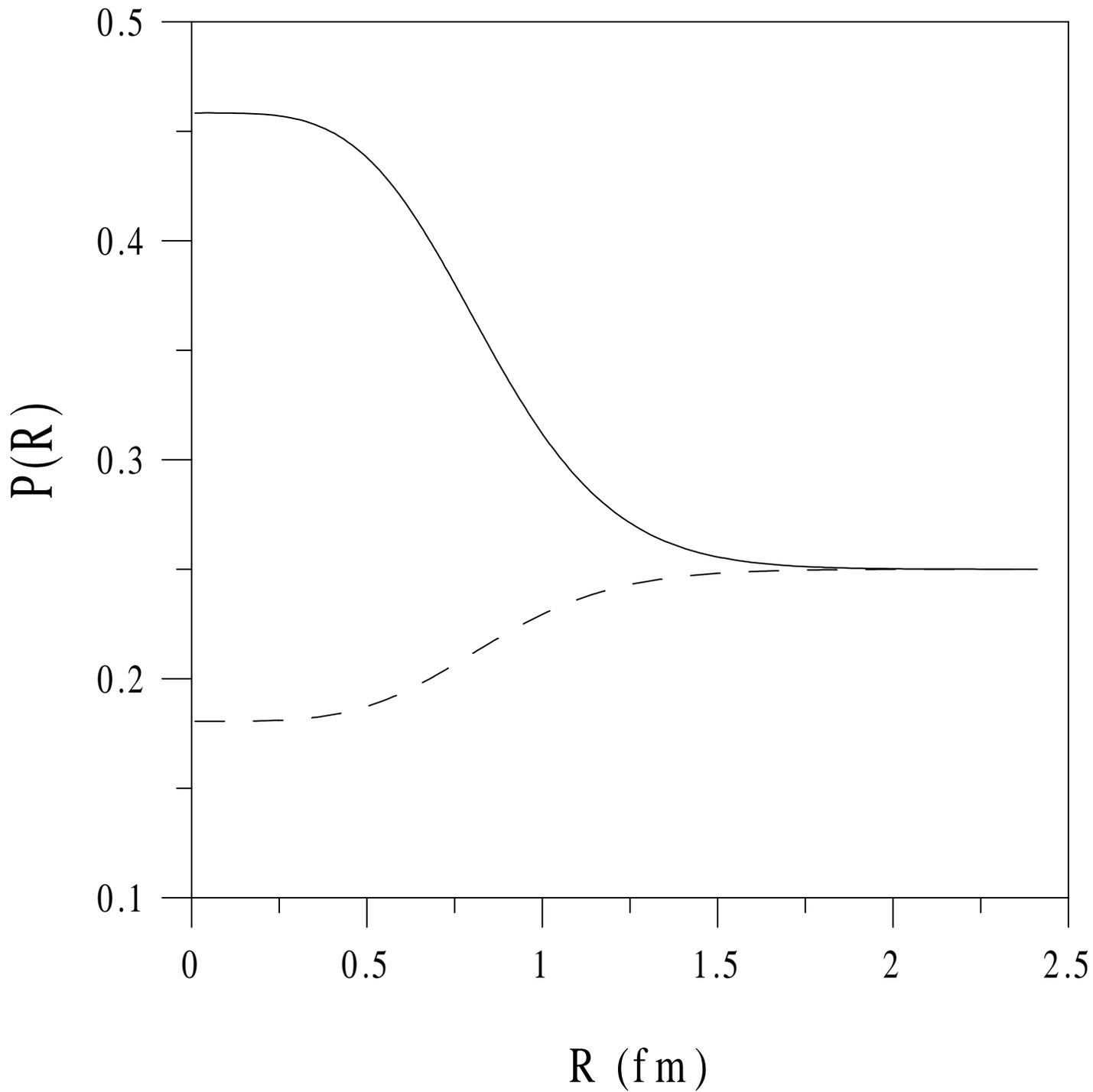
\begin{figure}[tbp]
\caption{Diquark-diquark probability in the $^1S_0$ $\Lambda\Lambda$ wave
function as a function of the interbaryon distance $R$. The solid line
represents the probability of the diquark-diquark component and the dashed
line the probability of any other non-diquark-diquark components.}
\label{fig4}
\end{figure}

On the theoretical side there has been also suggested that
inner diquark structures could prevent, through Pauli repulsion at
very short distances, such a tight binding \cite{Jaf04}. Then it may be
instructive to perform an analysis of the diquark contributions to our $%
\Lambda \Lambda $ interaction.
We can easily isolate components containing diquark-diquark structures from
the $\Lambda \Lambda $ wave function written above. In fact any $\Lambda $
wave function contains a diquark, say a pair of quarks in a $[1]_{{\rm spin}%
}-[\overline{3}]_{{\rm flavor}}-[\overline{3}]_{{\rm color}}$ configuration,
through the term $\phi ^{MA}\chi ^{MA}$ in equation (\ref{wfu}). The
contribution of these components to the $\Lambda \Lambda $ potential is
shown in Fig. \ref{fig3}. As can be checked it is attractive at all
distances and strongly attractive at short distances mainly due to the
gluonic Coulomb term and to a less extent to the $\sigma $ exchange. This
short-range attractive contribution (reduced by some repulsion from the rest
of the components) determines the short-range attractive character of the $%
\Lambda \Lambda $ potential. Actually the component containing
diquark-diquark becomes the dominant one when reducing the interbaryon
distance. This can be seen by defining a diquark-diquark probability
depending on the interbaryon distance $R$ as the ratio 
\begin{equation}
P(R)=\frac{\left\langle \left( \phi ^{MA}\chi ^{MA}\right) _{\Lambda }\left(
\phi ^{MA}\chi ^{MA}\right) _{\Lambda }\right\vert {\cal A}\left\vert \left(
\phi ^{MA}\chi ^{MA}\right) _{\Lambda }\left( \phi ^{MA}\chi ^{MA}\right)
_{\Lambda }\right\rangle }{4\left\langle \Phi _{\Lambda }\Phi _{\Lambda
}\right\vert {\cal A}\left\vert \Phi _{\Lambda }\Phi _{\Lambda
}\right\rangle }\,,
\end{equation}%
where we have used ${\cal A}^{2}={\cal A}$. This probability appears in Fig. %
\ref{fig4} where it is clear its increasing from the nonoverlapping value
0.25 to almost 0.5 at very short interbaryonic distances as it corresponds
to the bosonic diquark character (as a counterpart the other
non-diquark-diquark components show an opposite tendency). Therefore we do
not see in our analysis any signature of a strong Pauli diquark blocking at
short distances. This stems from the diquark-diquark structure we are
dealing with. Since both symmetric and antisymmetric color states are
allowed no restriction on the two identical diquark orbital angular momentum
comes out (quite a different situation one has in the hypothetical
pentaquark case where only the antisymmetric color state is allowed).
Therefore we conclude that the bound state found does not owe its loose
binding to Pauli diquark repulsion at very short distances but instead to
the diquark-diquark dynamics. If combined with the assumption that the
three-diquark component is dominant for the $H-$dibaryon such dynamics could
explain its non-appearance as a tightly bound state.

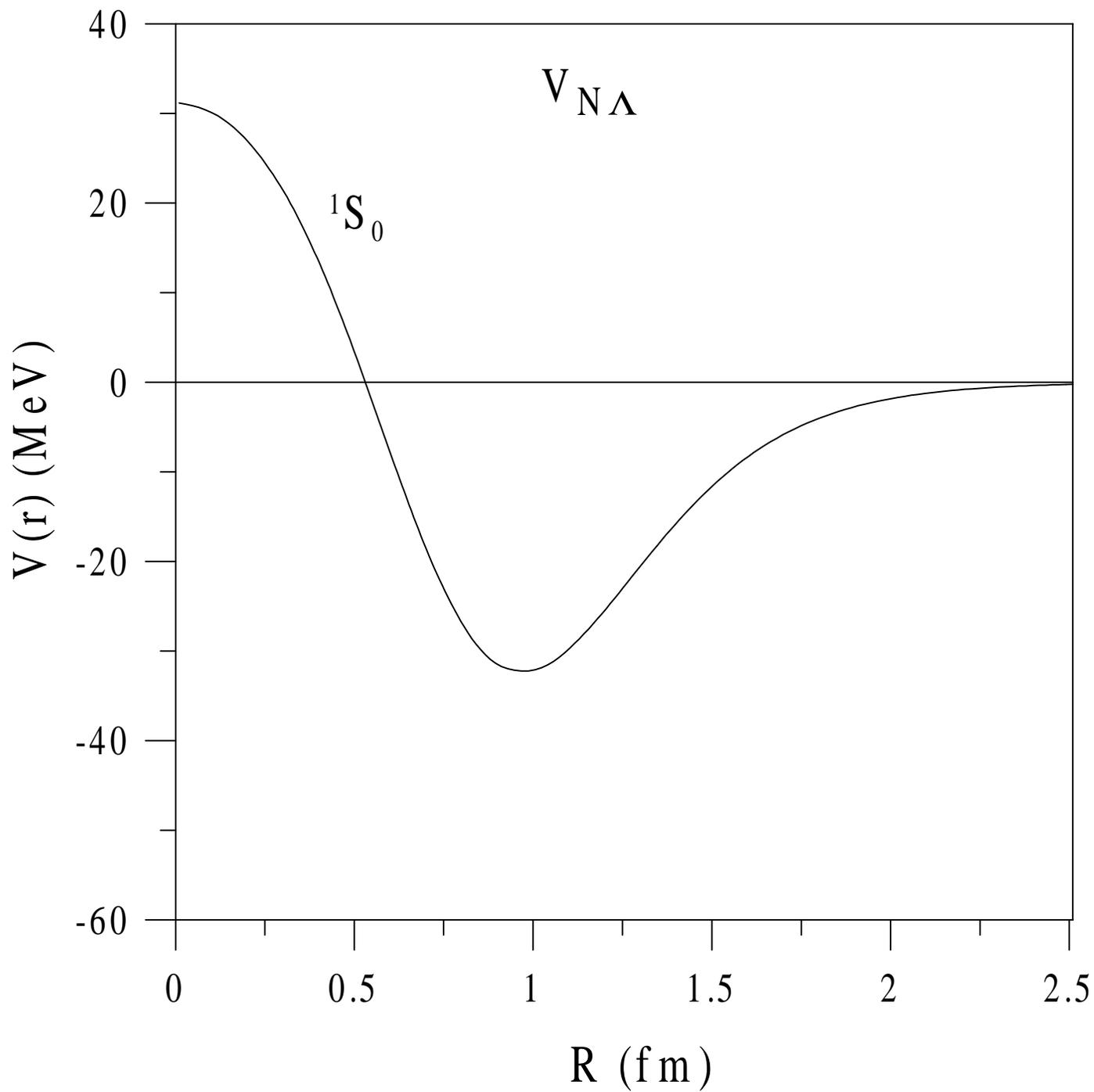
\begin{figure}[tbp]
\caption{$^1S_0$ $N \Lambda$ potential.}
\label{fig5}
\end{figure}

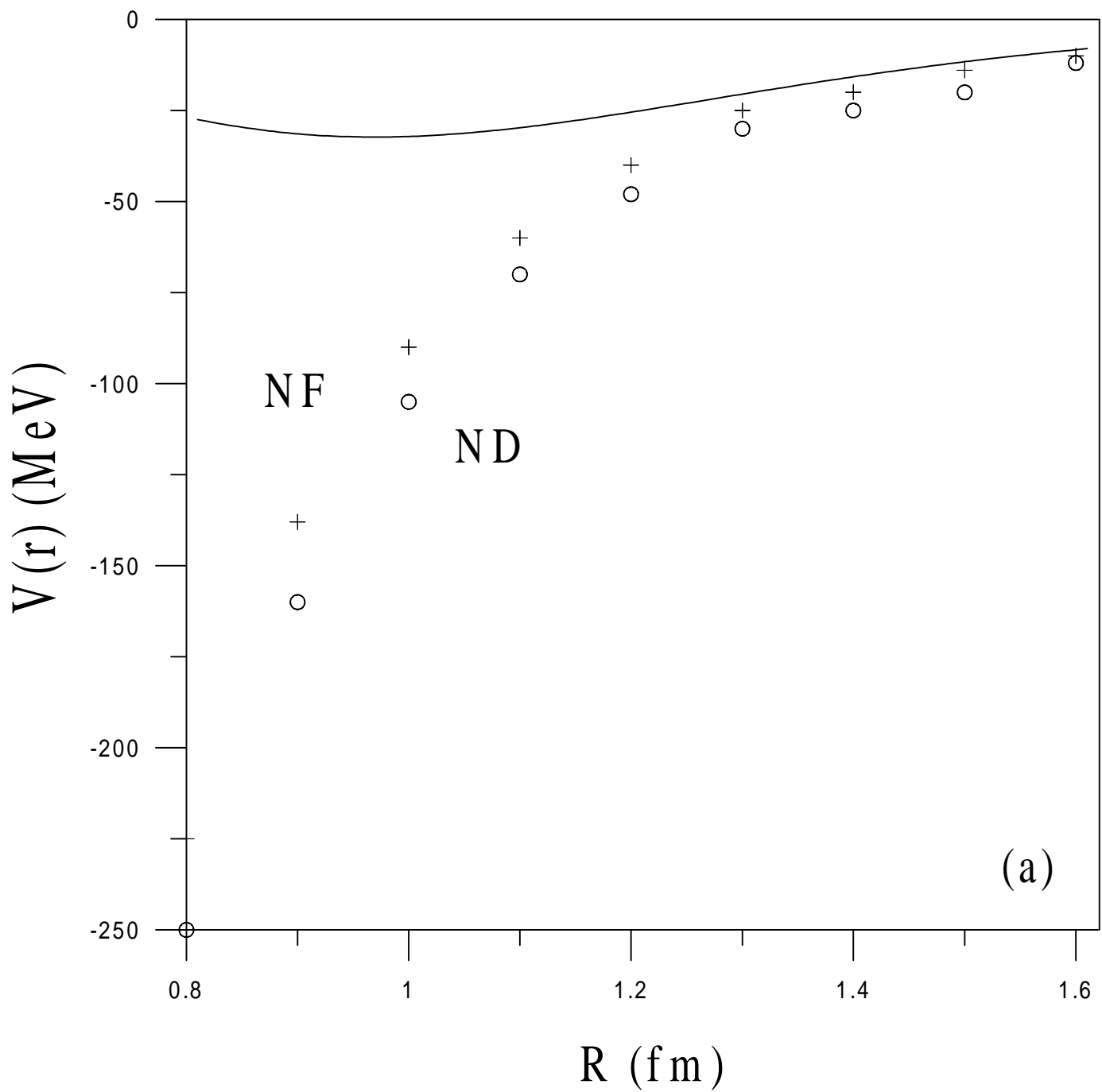
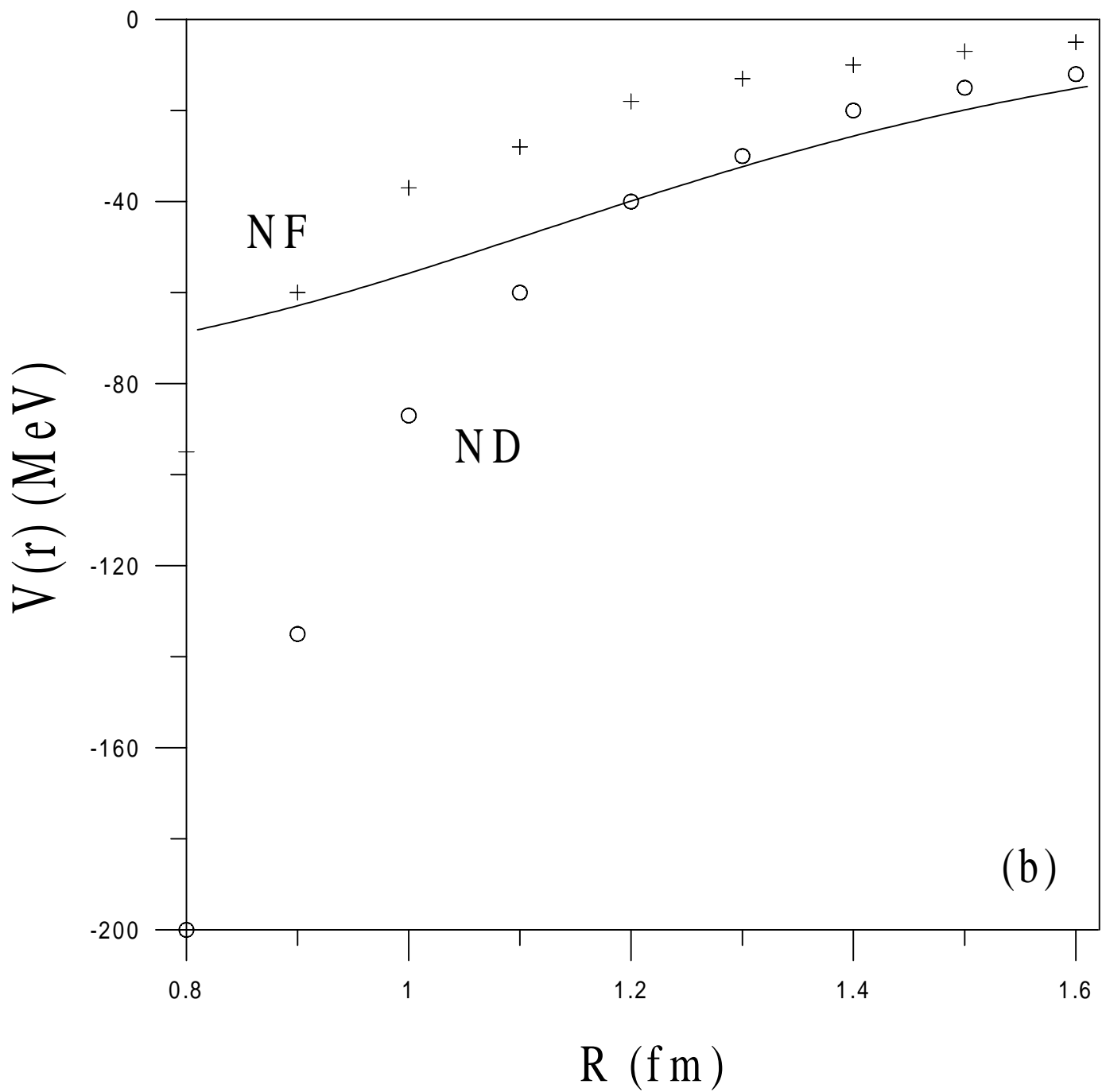
\begin{figure}[tbp]
\caption{(a) Comparison of the quark-model based $^1S_0$ $N \Lambda$
potential to the Nijmegen model (crosses stand for model F and circles for
model D) of Ref. \protect\cite{Yam94}. (b) Same as (a) for the $^1S_0$ $%
\Lambda\Lambda$ potential.}
\label{fig6}
\end{figure}

Finally, our results allow also for a direct determination of the $\sigma
\Lambda \Lambda $ coupling constant from the $\sigma qq$ one since at long
distances, where there is not any significant baryon overlap, the $\sigma $%
-exchange potential can be identified with a $\sigma $-exchange at the
baryon level that can be parametrized as, 
\begin{equation}
V^{\Lambda \Lambda \rightarrow \Lambda \Lambda }_{\sigma}(R)=-\frac{%
g_{\sigma \Lambda \Lambda }^{2}}{4\pi }\frac{\Lambda _{\chi}^{2}}{\Lambda
_{\chi}^{2}-m_{\sigma }^{2}}\frac{e^{-m_{\sigma }R}}{R} \, .
\end{equation}
In order to eliminate as much as possible the model dependence and to do a
non-meaningless comparison of our results with others in the literature it
is convenient to take the ratio to the $NN$ case \cite{Jul03}. Thus we get, 
\begin{equation}
\frac{g_{\sigma \Lambda \Lambda }}{g_{\sigma NN}}=0.88 \, .
\end{equation}
Let us emphasize that this ratio reflects precisely the different wave
function structure of $N$ and $\Lambda $ as a consequence of the $SU(3)$ quark
mass breaking. Our ratio is significantly bigger than the one predicted by
some baryonic models \cite{Hol95} but in agreement with others \cite%
{Reu94,Str88,Mae89,Hol89}. Since all baryonic models give a reasonable
description of the scarce $N\Lambda $ data one cannot discriminate between
them. This rather points out the need for more precise data. For the sake of
completeness we study the $N\Lambda $ potential with our model. The result
is shown in Fig. \ref{fig5}. As can be checked the interaction becomes
slightly repulsive at short distances due to the dominance of the pion plus
mainly gluon repulsion against the sigma attraction. Actually the pion and
gluon contributions are about twice those in $\Lambda \Lambda $ whereas the $%
\sigma $ contributions remain more or less the same. In Fig. \ref{fig6} we
compare our $N\Lambda $ potential with the Nijmegen interactions F and D 
\cite{Yam94}. As can be seen they are quite similar for $R\geq 1.3$ fm. In
the $\Lambda \Lambda $ case our results are definitely closer to the
predictions by Nijmegen interaction D.

In summary, by using a chiral constituent quark model precisely fitted
in the nonstrange sector to a bulk of baryon, meson and baryon-baryon data, we
have shown that, contrary to what has been usually assumed in baryonic
models and calculated in previous quark models, the $%
\Lambda \Lambda $ interaction in the $^{1}S_{0}$ partial wave is attractive
at short distances. In our CCM this attraction comes from the $\sigma $
exchange, whose effect cannot be surmounted even at short distances by the
gluon + pion + eta repulsion. The different short-range behavior as compared
to the $NN$ case is understood from the very different quantitative role
played by pion + gluon exchanges against sigma exchange in both cases.
Actually in the \textquotedblleft intermediate\textquotedblright\ $N\Lambda $
case our model interaction becomes repulsive at short distances in agreement
with previous treatments and suggestions from data. We predict a 
slightly bound $\Lambda \Lambda $ state whose energy fits well inside the
upper bounds imposed from $\Lambda \Lambda $ hypernuclei data. This
is encouraging to try to obtain a microscopic description of $\Lambda
\Lambda $ hypernuclei for which we have not at the current moment a
satisfactory explanation. To this respect to have at disposal 
the $\Lambda \Lambda$, $N\Lambda$ and  $NN$
interactions obtained on the same footing may be  extremely
valuable.

From the point of view of quark flavor-color configurations the $\Lambda
\Lambda $ attraction is mainly related to a structure containing two
diquarks. The involved diquark-diquark dynamics giving rise to the loosely
bound state could also provide an explanation for the experimental absence
of a tightly bound $H-$dibaryon. We have also derived a $g_{\sigma \Lambda
\Lambda }$ coupling being almost 90\% of $g_{\sigma NN}$ as a reflection of
the assumed $SU(3)$ breaking.

Certainly there are not at present precise data to check the $\Lambda
\Lambda $ short-range character and consequently our model against others.
Despite this we think it is worth to pursue a theoretical program aiming at
the examination of the consequences derived from it and their possible
experimental checks. This can be important not only to directly progress in
the understanding of the $\Lambda \Lambda $ system but also (due to the
connection of the short-range character to the gluon intensity) to
indirectly disentangle the quantitative role played by gluons and pions in
the $NN$ system.

\acknowledgements
This work was partially funded by Direcci\'{o}n General de Investigaci\'{o}n
Cient\'{\i}fica y T\'{e}cnica (DGICYT) under the Contract No. BFM2001-3563,
by Junta de Castilla y Le\'{o}n under the Contract No. SA104/04, and by
Oficina de Ciencia y Tecnolog\'\i a de la Comunidad Valenciana, Grupos03/094.

\begin{table}[tbp]
\caption{Quark-model parameters.}
\label{tabI}%
\begin{tabular}{cccc|ccc}
& $m_{u,d}$\,(MeV) & 313 &  & $\alpha_s $ & 0.54 &  \\ 
& $m_s$\,(MeV) & 555 &  & $r_0$\,(fm$^{-1}$) & 0.18 &  \\ 
& $m_{\pi}$\,(fm$^{-1}$) & 0.7 &  & $a_c$\,(MeV\,fm$^{-1}$) & 185.0 &  \\ 
& $m_\sigma$\,(fm$^{-1}$) & 3.42 &  & ${g^2_{ch}/{4\pi}}$ & 0.54 &  \\ 
& $m_K$\,(fm$^{-1}$) & 2.51 &  & $\Lambda_{\chi}$\,(fm$^{-1}$) & 4.2 &  \\ 
& $m_{\eta}$\,(fm$^{-1}$) & 2.77 &  & $\Lambda_{K}= \Lambda_{\eta}$\,(fm$%
^{-1}$) & 5.2 &  \\ 
& $\theta_p \,(^o)$ & $-$15 &  &  &  &  \\ 
&  &  &  &  &  & 
\end{tabular}%
\end{table}

\end{document}